\documentclass[11pt]{article}
\usepackage{authblk}
\usepackage{amsmath,amssymb,mathrsfs,centernot,amsfonts}
\usepackage{enumitem}
\usepackage{verbatim}
\usepackage{color}
\usepackage[margin=1in]{geometry}
\usepackage{setspace}
\usepackage{graphics,graphicx}
\usepackage{multicol}

\newcommand{\lang}{\mathcal{L}}
\newcommand{\limp}{\longrightarrow}
\newcommand{\pow}{\mathcal{P}}
\newcommand{\V}{\mathscr{V}}
\newcommand{\F}{\mathscr{F}}
\newcommand{\var}{\mathsf{Var}}
\newcommand{\mone}{\models_{\mathrm{UKN1}}}
\newcommand{\mtwo}{\models_{\mathrm{UKN2}}}

\usepackage{amsthm}
\theoremstyle{definition}

\usepackage[backref=page,colorlinks=true,allcolors=blue]{hyperref}

  \title{How a computer might think}
  \author[1]{Sankha S. Basu}
  \date{June 21, 2026}
  \affil[1]{Department of Mathematics\\
  Indraprastha Institute of Information Technology-Delhi\\
  New Delhi, India.}

\begin{document}

\maketitle

\begin{abstract}
    Inspired by Nuel Belnap's "How a computer should think," which gave rise to the four-valued logic FDE, we contemplate, in this article, how a computer might think if we add a fifth value for unknowable or cannot be known. We devise two new five-valued logics, UKN1 and UKN2, called the logics of the unknowable. These are different from the five-valued logic FDEe of the FDE-family. The main difference is in the number of designated truth values. While FDEe takes two designated values, UKN1 and UKN2 have three. The four-valued reducts of these logics are also different from FDE. This is due to the fact that  instead of taking one of the non-designated values as neither true nor false, as in FDE, we have interpreted this as not known yet. This value although denoted by the same letter $n$, behaves differently from its namesake in FDE.
\end{abstract}

\textbf{Keywords:}
Five-valued logic; many-valued logic; unknowable

\section{Introduction}

Nuel Belnap in \cite{Belnap1977-1} (reprinted as \cite{Belnap2019-1}) envisions how a computer should think taking into consideration potential inconsistencies and incompleteness in its received information. This results in a four-valued logic in \cite{Belnap1977-2} (reprinted as \cite{Belnap2019-2}), called the logic of \emph{first degree entailment} (FDE), which is inconsistency tolerant, and more specifically, \emph{paraconsistent}. A paraconsistent logic is one that fails the law of explosion, i.e., the classical principle which dictates that everything follows from a contradiction. Now, this is of course quite useful for a modern computer which might be subjected to inconsistent information arising due to various possible reasons, such as multiple sourcing of the  data, or error on the part of the data entry operator. Ever since its introduction, FDE has been heavily studied and now has acquired a substantial legacy. This is outlined in \cite{OmoriWansing2017}.

Thomas Ferguson in \cite{Ferguson2016} considers the situation in which a Belnap computer is unable to answer a query or encounters a fault. In this article the semantic values of FDE are represented as data pairs corresponding to a truth value and a falsity value. Then, depending on whether these values are stored at a single address or two distinct addresses, two five-valued logics emerge. While the single address variant leads to FDE$_\varphi$, a logic introduced by Graham Priest in \cite{Priest2010} as a formalization of later Buddhist thought, the two address alternative is found to be the logic of \emph{analytic containment} (AC) introduced by Richard Angell in \cite{Angell1977}.

In this connection it must be mentioned that it has been argued by Priest in several places, including \cite{Priest2010,Priest2022}, that FDE is the formalization of the original four-cornered system of Buddhist logic, the four corners representing the four possible answers to an assertion/denial query, viz., yes, no, both, and neither. These correspond to the four semantic values in FDE, \emph{true, false, both}, and \emph{neither}. Matters become a bit more complicated in later Buddhist thought, where a fifth possibility emerges, viz., none of the four. This state is supposed to represent the silence of Buddha on certain questions. Priest calls this \emph{ineffability} and formulates the five-valued logic FDE$_\varphi$ in \cite{Priest2010} or FDEe in \cite{Priest2019,Priest2022}. There is, however, a debate about whether the `none of the four corners' or ineffability should be taken in the same stride as the original four values. We do not go into this debate here and, in fact, we steer clear of the question of whether FDE and FDEe are apt formalizations of Buddhist thought.

We return to the issue of how a computer, or a `canon of inference' as Belnap calls it, might work. We propose slightly different machines that can, at least in some instances, distinguish between `not known (yet)' and `cannot be known' or `unknowable.' While the not known indicates that there is a possibility of knowing in the future or as the computation progresses, the unknowable indicates that there is no such hope. Thus, there is a sense of openness, as in open to the possibility of a change in status, when the answer is `not known.' On the other hand, there is a sense of finality or closure in case the answer is `cannot be known.' We will thus refer to `not known' as an \emph{open} semantic state or open value, and `cannot be known' as a \emph{closed} semantic state or closed value. Such machines would thus be able to respond to an assertion/denial query with five possible responses: true, false, both, not known, and unknowable. The three semantic values true, false and both, i.e., simultaneous assertion and denial are regarded as closed. We will assume that the computer stops once it encounters either of these values.

One other variation of the above would be if we treat the `unknowable' state to indicate that there is a fault in the system and due to that the unknowable status has been reached. In this particular situation, this state would still be considered closed but become infectious as we will argue and the logic would be different from the system described above. Finally, just to round things off, let's consider the situation when all values are considered closed. Then, it is not hard to see that the unknowable and not known collapse onto each other and we get a four-valued system. 

In what follows, two new five-valued systems are discussed. These share similarities with FDE and FDEe, which will be pointed out. The new logics may be called the logics of the unknowable and denoted by UKN1 and UKN2.

\section{Preliminaries}\label{sec:prelim}
Let $\Sigma$ be a finite set of connectives or operators, called the \emph{language} or \emph{signature}, and $\var$ be a denumerable set of propositional variables or atoms. In this paper, we always assume that $\Sigma=\{\neg,\land,\lor\}$. We will denote the elements of $\var$ by $p,q,r$, etc. The set of formulas $\lang$ is then generated inductively over $\var$ using the elements of the signature, as usual.

A \emph{logical matrix} or \emph{matrix} is an ordered triple $\langle\V,\F,D\rangle$, where
    \begin{enumerate}[label=(\roman*)]
        \item $\V$ is the set of \emph{truth values}.
        \item $\F=\{f_\neg:\V\to\V,f_\land:\V^2\to\V,f_\lor:\V^2\to\V\}$, the set of three functions corresponding to the three connectives in $\Sigma$.
        \item $D\subsetneq\V$ is the set of \emph{designated values}.
    \end{enumerate}

    Given a matrix $\langle\V,\F,D\rangle$, a \emph{valuation} is a function $v:\var\to\V$. This can then be extended to a unique function $V:\lang\to\V$ such that
    \begin{enumerate}[label=(\roman*)]
        \item $V(p)=v(p)$ for all $p\in \var$,
        \item $V(\neg\alpha)=f_\neg(V(\alpha))$ for all $\alpha\in\lang$.
        \item $V(\alpha_1\circ\alpha_2)=f_\circ(V(\alpha_1),V(\alpha_2))$ for all $\alpha_1,\alpha_2\in\lang$, where $\circ\in\{\land,\lor\}$.
    \end{enumerate}

Since $v$ and $V$ determine each other uniquely, we will drop the distinction between the two and use the notation $v$ for both valuations and their extensions.
    
Finally, given a matrix $\langle\V,\F,D\rangle$, the \emph{logic} (induced by the matrix) $\mathscr{L}$ is a pair $\langle\lang,\models_\mathscr{L}\rangle$, where $\models_\mathscr{L}\,\subseteq\pow(\lang)\times\lang$ ($\pow(\lang)$ denotes the power set of $\lang$) is the \emph{consequence} relation, defined as follows. For any $\Gamma\cup\{\alpha\}\subseteq\lang$, 
    \[
    \Gamma\models_\mathscr{L}\alpha\;\hbox{ iff }\;v(\Gamma)\subseteq D\hbox{ implies }v(\alpha)\in D,\hbox{ for every valuation }v.
    \]

We drop the subscript $\mathscr{L}$ from $\models_\mathscr{L}$ only when there is no chance of confusion. For the logics to follow in the next sections, we will have $\V=\{t,f,b,n,u\}$ as the set of truth values, where 
$t,f,b,n,u$ represent true, false, both, not known yet, and unknowable, respectively. We will take the set of designated values as $D=\{t,b,u\}$. The justifications for this choice will be discussed further in the following sections.

\section{The logic UKN1}
We start, as indicated above, with the set of formulas $\lang$ generated inductively over a denumerable set of propositional variables $\var$ using the signature $\Sigma=\{\neg,\land,\lor\}$. The set of truth values $\V=\{t,f,b,n,u\}$, where $t,f,b,n,u$ are as mentioned above. We consider a computer that stops if the output is one of $t,f,b$, or $u$. On the other hand, if the output is $n$, then that indicates an ongoing computation and that the value is not known yet. We might as well see these as states of the computer when a particular query is made. Thus, we will use the terms truth values and truth states interchangeably. As discussed in the introductory section, we will call the values $t,f,b$, and $u$ \emph{closed}, while the truth state $n$ will be referred to as \emph{open}. The underlying intuition is that an open state indicates a computation in progress and as the computation progresses, the computer might find itself with one of the closed states where it halts. The closed and non-false values are taken as designated. Thus, $D=\{t,b,u\}$. Now, the last and most crucial ingredient of the logical matrix is $\F=\{f_\neg,f_\land,f_\lor\}$. Let's take the case of $f_\neg$ first. Negation should turn $t$ to $f$ and vice versa. If a statement is either both true and false, or unknowable, or not known yet, then so is their negation. Thus the negation leaves the values $b,n$, and $u$ unchanged. This gives us the following definition for $f_\neg:\V\to\V$, which is the truth table for negation.
\[
\begin{array}{c|c|c|c|c|c}
     &t&f&b&n&u\\
     \hline
     f_\neg&f&t&b&n&u
\end{array}
\]
Next, we take up the case of conjunction, i.e., the definition of $f_\land:\V^2\to\V$. The values $t$ and $f$, of course, behave classically. The behavior of $b$ with respect to $t,f$, and $b$ is as in FDE. Thus, we get the following partial table for conjunction.
\[
\begin{array}{c|ccccc}
     f_\land&t&f&b&n&u\\
     \hline
     t&t&f&b\\
     f&f&f&f\\
     b&b&f&b\\
     n&\\
     u&
\end{array}
\]
The justifications for these are quite straightforward. For example, suppose we have a statement that is true and another that is both true and false. Since both statements are true, the conjunction is true. However, since one of them is also false, the conjunction is also false. Thus, the conjunction must be both true and false. The rest of the values in the above table can be found via analogous reasoning.

Things become interesting when the values $n$ and $u$ come into play. Let's look into the case of $u$ first. Suppose we have a statement whose truth value is unknowable. Then its conjunction with a statement whose truth value is either $t$, $b$ or $u$ is also unknowable. However, its conjunction with a false statement is false on account of the one false conjunct.

Lastly, suppose a statement has the value $n$. Then as per our design, the truth state of the statement might later acquire one of the closed values $t,f,b$, or $u$. Now, if we are to find the truth value of the conjunction of this statement with a statement for which the computer has yielded the value $t$, then we ought to consider all the possible future values of the conjunction. If the future value of the statement with current value $n$ is $t,f,b$, or $u$, then the conjunction with a true statement will land, respectively, on the values $t$, $f$, $b$ and $u$. Thus, the value of the conjunction is also not known yet, i.e., $n$. A similar reasoning yields the observation that the value of a conjunction of two statements, one with value $n$ and the other with value $b$ or $u$ should be $n$. However, if the statement with current value $n$ enters into a conjunction with a false statement, then no matter what the future value of the statement is, the conjunction will be false and hence, in this case the conjunction will have value $f$. Thus, we get the following table for conjunction.

\[
\begin{array}{c|ccccc}
     f_\land&t&f&b&n&u\\
     \hline
     t&t&f&b&n&u\\
     f&f&f&f&f&f\\
     b&b&f&b&n&u\\
     n&n&f&n&n&n\\
     u&u&f&u&n&u
\end{array}
\]

Now, contemplating on similar lines but this time for disjunction, we get the following table for $f_\lor:\V^2\to\V$.

\[
\begin{array}{c|ccccc}
     f_\lor&t&f&b&n&u\\
     \hline
     t&t&t&t&t&t\\
     f&t&f&b&n&u\\
     b&t&b&b&n&u\\
     n&t&n&n&n&n\\
     u&t&u&u&n&u
\end{array}
\]

In this case, we see that it is sufficient for a disjunction to be true if one of the disjuncts is so. As a further example, let's take the case of the disjunction of two statements, one with value $n$ and the other with value $u$. The statement with value $n$ might later acquire one of the values $t,f,b$, or $u$. So, then the disjunction will get the value $t$ or $u$. Thus, presently, the value of the disjunction has to be marked as $n$.

One case that is not discussed above is what happens when the value of a statement with current state $n$ continues to be at that state and does not take any of the closed values. This does not create any problems in the setup as the value of the conjunction or disjunction of it with any other statement can still be computed based on the tables above. For example, the disjunction of such a statement with a statement with value $b$ will continue to be at $n$.

It is clear from the above tables that the values $t,f$, and $b$ behave exactly the same way among themselves as in FDE or FDEe. The value $n$ behaves in a different way from that in FDE or FDEe. The truth value of the conjunction of statements with values $n$ and $b$ is not $f$ as is the case in FDE or FDEe. Nor is it that the disjunction of such statements takes the value $t$. As a result of this, the diamond lattice structure of the FDE truth values is not found for UKN1.

The logic FDEe uses the same set of designated values as FDE, i.e., the fifth truth value $e$ is not designated here. UKN1 on the other hand, has three designated values. Moreover, the fifth truth value $e$ of FDEe is \emph{infectious}, i.e., a statement gets the value $e$ iff one of its components has the value $e$. UKN1 differs from FDEe there as well, since none of the values is infectious.

We thus have finished the construction of the matrix for UKN1. Next, we define UKN1-valuations and the UKN1 consequence relation $\mone$ as per the recipe laid out in Section \ref{sec:prelim}. Thus, UKN1 is the logic induced by the matrix $\langle\V,\F,D\rangle$, such that $\V=\{t,f,b,n,u\}$, $D=\{t,b,u\}$ and $\F=\{f_\neg,f_\land,f_\lor\}$, where $f_\neg,f_\land$ and $f_\lor$ are given by the tables above. 

Instead of using the above truth tables, the UKN1-valuations can be alternatively described in a linear form as follows. Suppose $\alpha,\beta\in\lang$ and $v$ is a UKN1-valuation.

\textsc{Negation}
\begin{itemize}
    \item $v(\neg\alpha)=t$ iff $v(\alpha)=f$
    \item $v(\neg\alpha)=f$ iff $v(\alpha)=t$
    \item $v(\neg\alpha)=b$ iff $v(\alpha)=b$
    \item $v(\neg\alpha)=n$ iff $v(\alpha)=n$
    \item $v(\neg\alpha)=u$ iff $v(\alpha)=u$
\end{itemize}

\textsc{Conjunction}
\begin{itemize}
    \item $v(\alpha\land\beta)=t$ iff $v(\alpha)=t$ and $v(\beta)=t$
    \item $v(\alpha\land\beta)=f$ iff $v(\alpha)=f$ or $v(\beta)=f$
    \item $v(\alpha\land\beta)=b$ iff ($v(\alpha)=b$ and $v(\beta)=t/b$) or ($v(\beta)=b$ and $v(\alpha)=t/b$) 
    \item $v(\alpha\land\beta)=n$ iff ($v(\alpha)=n$ and $v(\beta)\neq f$) or ($v(\beta)=n$ and $v(\alpha)\neq f$)
    \item $v(\alpha\land\beta)=u$ iff ($v(\alpha)=u$ and $v(\beta)\neq n/f$) or ($v(\beta)=u$ and $v(\alpha)\neq n/f$)
\end{itemize}

\textsc{Disjunction}
\begin{itemize}
    \item $v(\alpha\lor\beta)=t$ iff $v(\alpha)=t$ or $v(\beta)=t$
    \item $v(\alpha\lor\beta)=f$ iff $v(\alpha)=f$ and $v(\beta)=f$
    \item $v(\alpha\lor\beta)=b$ iff ($v(\alpha)=b$ and $v(\beta)=f/b$) or ($v(\beta)=b$ and $v(\alpha)=f/b$) 
    \item $v(\alpha\lor\beta)=n$ iff ($v(\alpha)=n$ and $v(\beta)\neq t$) or ($v(\beta)=n$ and $v(\alpha)\neq t$)
    \item $v(\alpha\lor\beta)=u$ iff ($v(\alpha)=u$ and $v(\beta)\neq n/t$) or ($v(\beta)=u$ and $v(\alpha)\neq n/t$)
\end{itemize}

We now investigate the properties of the logic UKN1 $=\langle\lang,\mone\rangle$. The first observation is that there is no formula $\alpha\in\lang$ such that $v(\alpha)\in D$ for all valuations $v$. This can be seen by noting from the tables above that for any valuation $v$, such that $v(p)=n$ for each variable $p$ occurring in $\alpha$, $v(\alpha)=n\in\V\setminus D$. Thus, UKN1 has no tautologies. This is a property that it shares with FDE \cite{OmoriWansing2017}. In particular, this implies that there exists $\alpha$ such that $\not\mone\alpha\lor\neg\alpha$ and $\not\mone\neg(\alpha\land\neg\alpha)$, i.e., the \emph{law of excluded middle (LEM)} and the \emph{law of non-contradiction (LNC)} fail in UKN1. The failure of LEM indicates that the logic is \emph{paracomplete}. 

Now, suppose $p,q\in\var$ are distinct and $v$ is a UKN1-valuation such that $v(p)=b$ and $v(q)=n$. Then, $v(\{p,\neg p\})=\{b\}\subseteq D$, while $v(q)\notin D$. This shows that $\{p,\neg p\}\not\mone q$. (The same result follows by taking $v(p)=u$ and $v(q)=n$ or $f$.) Thus, the \emph{law of explosion} or \emph{ex contradictione quodlibet (ECQ)} fails in UKN1, which means that the logic is \emph{paraconsistent}. This, coupled with the failure of the LNC, implies that UKN1 is, in fact, \emph{strongly} paraconsistent.

We mention some notable failures of the UKN1-consequence before going into positive results. The first of these is the standard or classical rule of $\lor$-introduction, which can be expressed as $\alpha\models\alpha\lor\beta$, for any two formulas $\alpha,\beta$. This can be seen to fail in UKN1 by considering distinct $p,q\in\var$ and a valuation $v$ such that $v(p)=b$ (or $v(p)=u$) and $v(q)=n$. Then, $v(p\lor q)=n$. Thus, $v(p)\in D$, while $v(p\lor q)\notin D$, which implies that $p\not\mone p\lor q$.

However, the standard disjunction elimination rule holds in UKN1. Thus, for any $\alpha,\beta,\gamma\in\lang$, 
\[
\hbox{if }\alpha\mone\gamma\hbox{ and }\beta\mone\gamma,\hbox{ then }\alpha\lor\beta\mone\gamma.
\]
\begin{proof}
    Suppose $\alpha\mone\gamma$ and $\beta\mone\gamma$ but $\alpha\lor\beta\not\mone\gamma$. Then, there exists a UKN1-valuation $v$ such that $v(\alpha\lor\beta)\in D$ but $v(\gamma)\notin D$. Now, since $\alpha\mone\gamma$ and $\beta\mone\gamma$, $v(\gamma)\notin D$ implies that $v(\alpha),v(\beta)\notin D$. However, that implies that $v(\alpha\lor\beta)\notin D$, which is a contradiction. Hence, the result follows.
\end{proof}

The distributive laws fail. There exist $\alpha,\beta,\gamma\in\lang$ such that
\begin{enumerate}[label=(\roman*)]
    \item $\alpha\land(\beta\lor\gamma)\not\mone(\alpha\land\beta)\lor(\alpha\land\gamma)$; and
    \item $\alpha\lor(\beta\land\gamma)\not\mone(\alpha\lor\beta)\land(\alpha\lor\gamma)$.
\end{enumerate}

\begin{proof}
    \begin{enumerate}[label=(\roman*)]
        \item Let $p,q,r\in\var$ be distinct and $v$ be a UKN1-valuation such that $v(p)=b,v(q)=n$, and $v(r)=t$. Then, $v(q\lor r)=t$, and hence, $v(p\land(q\lor r))=b\in D$. On the other hand, $v(p\land q)=n$ and $v(p\land r)=b$. Thus, $v((p\land q)\lor(p\land r))=n\notin D$. Hence, $p\land(q\lor r)\not\mone(p\land q)\lor(p\land r)$.

        \item Let $p,q,r\in\var$ be distinct and $v$ be a UKN1-valuation such that $v(p)=b,v(q)=n$, and $v(r)=f$. Then, $v(q\land r)=f$, and hence, $v(p\lor(q\land r))=b\in D$. On the other hand, $v(p\lor q)=n$ and $v(p\lor r)=b$. Thus, $v((p\lor q)\land(p\lor r))=n\notin D$. Hence, $p\lor(q\land r)\not\mone(p\lor q)\land(p\lor r)$.
    \end{enumerate}
\end{proof}

However, for any $\alpha,\beta,\gamma\in\lang$, we have the following entailments.
\begin{enumerate}[label=(\roman*)]
    \item $(\alpha\land\beta)\lor(\alpha\land\gamma)\mone\alpha\land(\beta\lor\gamma)$; and
    \item $(\alpha\lor\beta)\land(\alpha\lor\gamma)\mone\alpha\lor(\beta\land\gamma)$.
\end{enumerate}

\begin{proof}
    \begin{enumerate}[label=(\roman*)]
        \item Suppose $v$ is a UKN1-valuation such that $v(\alpha\land(\beta\lor\gamma))\notin D$.

        \textsc{Case 1:} $v(\alpha\land(\beta\lor\gamma))=f$.

        This implies that either $v(\alpha)=f$ or $v(\beta\lor\gamma)=f$. If $v(\alpha)=f$, then $v(\alpha\land\beta)=v(\alpha\land\gamma)=f$, which leads to $v((\alpha\land\beta)\lor(\alpha\land\gamma))=f$. On the other hand, if $v(\beta\lor\gamma)=f$, then $v(\beta)=v(\gamma)=f$. This implies that $v(\alpha\land\beta)=v(\alpha\land\gamma)=f$, and thus, $v((\alpha\land\beta)\lor(\alpha\land\gamma))=f$.

        \textsc{Case 2:} $v(\alpha\land(\beta\lor\gamma))=n$.

        This implies that either $v(\alpha)=n$ and $v(\beta\lor\gamma)\neq f$, or $v(\beta\lor\gamma)=n$ and $v(\alpha)\neq f$.

        In the first scenario, when $v(\alpha)=n$, $v(\alpha\land\beta)=n/f$, and similarly, $v(\alpha\land\gamma)=n/f$. Thus, $v((\alpha\land\beta)\lor(\alpha\land\gamma))=n$.

        On the other hand, when $v(\beta\lor\gamma)=n$ and $v(\alpha)\neq f$, we have either $v(\beta)=n$ and $v(\gamma)\neq t$, or $v(\gamma)=n$ and $v(\beta)\neq t$. Suppose $v(\beta)=n$ and $v(\gamma)\neq t$. Then, as $v(\alpha)\neq f$, $v(\alpha\land\beta)=n$, and since $v(\gamma)\neq t$, $v(\alpha\land\gamma)\neq t$. Thus, $v((\alpha\land\beta)\lor(\alpha\land\gamma))=n$. The case where $v(\gamma)=n$ and $v(\beta)\neq t$ is similar and also yields the same outcome, i.e., $v((\alpha\land\beta)\lor(\alpha\land\gamma))=n$.

        Thus, in all cases, $v((\alpha\land\beta)\lor(\alpha\land\gamma))\notin D$. Hence, by contraposition, for any UKN1-valuation $v$, if $v((\alpha\land\beta)\lor(\alpha\land\gamma))\in D$, then $v(\alpha\land(\beta\lor\gamma))\in D$. So, $(\alpha\land\beta)\lor(\alpha\land\gamma)\mone\alpha\land(\beta\lor\gamma)$.

        \item Suppose $v$ is a UKN1-valuation such that $v(\alpha\lor(\beta\land\gamma))\notin D$.

        \textsc{Case 1:} $v(\alpha\lor(\beta\land\gamma))=f$.

        So, $v(\alpha)=f$ and $v(\beta\land\gamma)=f$. Now, $v(\beta\land\gamma)=f$ implies that either $v(\beta)=f$ or $v(\gamma)=f$. If $v(\beta)=f$, then $v(\alpha\lor\beta)=f$ and so, $v((\alpha\lor\beta)\land(\alpha\lor\gamma))=f$. If on the other hand, $v(\gamma)=f$, then $v(\alpha\lor\gamma)=f$, and we have the same outcome, i.e., $v((\alpha\lor\beta)\land(\alpha\lor\gamma))=f$.

        \textsc{Case 2:} $v(\alpha\lor(\beta\land\gamma))=n$.

        In this case, either $v(\alpha)=n$ and $v(\beta\land\gamma)\neq t$, or $v(\beta\land\gamma)=n$ and $v(\alpha)\neq t$.

        Suppose first that $v(\alpha)=n$ and $v(\beta\land\gamma)\neq t$. Since, $v(\beta\land\gamma)\neq t$, either $v(\beta)\neq t$ or $v(\gamma)\neq t$. If $v(\beta)\neq t$, then $v(\alpha\lor\beta)=n$, and hence, $v((\alpha\lor\beta)\land(\alpha\lor\gamma))=n/f$. Similarly, if $v(\gamma)\neq t$, then $v(\alpha\lor\gamma)=n$, and so, $v((\alpha\lor\beta)\land(\alpha\lor\gamma))=n/f$.

        Now, suppose $v(\beta\land\gamma)=n$ and $v(\alpha)\neq t$. We note that $v(\beta\land\gamma)=n$ implies that either $v(\beta)=n$ and $v(\gamma)\neq f$, or $v(\gamma)=n$ and $v(\beta)\neq f$. If $v(\beta)=n$, then as $v(\alpha)\neq t$, $v(\alpha\lor\beta)=n$, and hence, $v((\alpha\lor\beta)\land(\alpha\lor\gamma))=n/f$. Similarly, if $v(\gamma)=n$, then $v(\alpha\lor\gamma)=n$, and hence, $v((\alpha\lor\beta)\land(\alpha\lor\gamma))=n/f$.

        Thus, in all cases, $v((\alpha\lor\beta)\land(\alpha\lor\gamma))\notin D$. Hence, by contraposition, for any UKN1-valuation $v$, if $v((\alpha\lor\beta)\land(\alpha\lor\gamma))\in D$, then $v(\alpha\lor(\beta\land\gamma))\in D$. So, $v((\alpha\lor\beta)\land(\alpha\lor\gamma))\mone\alpha\lor(\beta\land\gamma)$.
    \end{enumerate}
\end{proof}

Thus, we have one-sided distributive laws in UKN1. We now list some other valid entailments in UKN1. These can be proved using the  truth tables. We will prove some to exemplify. The rest can be argued for similarly. Suppose $\alpha,\beta,\gamma\in\lang$.

\textsc{Double negation laws:}
\begin{multicols}{2}
\begin{enumerate}[label=(\roman*)]
    \item $\neg\neg\alpha\mone\alpha$
    \item $\alpha\mone\neg\neg\alpha$
\end{enumerate}
\end{multicols}

\textsc{Commutative laws:}
\begin{multicols}{2}
    \begin{enumerate}[label=(\roman*),start=3]
        \item $\alpha\land\beta\mone\beta\land\alpha$
        \item $\alpha\lor\beta\mone\beta\lor\alpha$
    \end{enumerate}
\end{multicols}

\textsc{Conjunction laws:}
\begin{multicols}{2}
    \begin{enumerate}[label=(\roman*),start=5]
        \item $\alpha\land\beta\mone\alpha$
        \item $\{\alpha,\beta\}\mone\alpha\land\beta$
    \end{enumerate}
\end{multicols}

\textsc{Associative laws:}
\begin{multicols}{2}
\begin{enumerate}[label=(\roman*),start=7]
    \item $\alpha\land(\beta\land\gamma)\mone(\alpha\land\beta)\land\gamma$
    \item $(\alpha\land\beta)\land\gamma\mone\alpha\land(\beta\land\gamma)$
\end{enumerate}
\end{multicols}

\begin{multicols}{2}
\begin{enumerate}[label=(\roman*),start=9]
    \item $\alpha\lor(\beta\lor\gamma)\mone(\alpha\lor\beta)\lor\gamma$
    \item $(\alpha\lor\beta)\lor\gamma\mone\alpha\lor(\beta\lor\gamma)$
\end{enumerate}
\end{multicols}

\textsc{De Morgan's laws:}
\begin{multicols}{2}
    \begin{enumerate}[label=(\roman*),start=11]
        \item $\neg(\alpha\land\beta)\mone\neg\alpha\lor\neg\beta$
        \item $\neg\alpha\lor\neg\beta\mone\neg(\alpha\land\beta)$
    \end{enumerate}
\end{multicols}

\begin{multicols}{2}
    \begin{enumerate}[label=(\roman*),start=13]
        \item $\neg(\alpha\lor\beta)\mone\neg\alpha\land\neg\beta$
        \item $\neg\alpha\land\neg\beta\mone\neg(\alpha\lor\beta)$
    \end{enumerate}
\end{multicols}

\begin{proof}
    The proofs of (i)--(vi) are can be read off easily from the truth tables listed earlier. We prove (viii), (xi), and (xiv) below. The rest can be proved similarly by analyzing the truth tables and following similar argument patterns.

    \begin{enumerate}[label=(\roman*),start=8]
        \item Suppose $v$ is a UKN1-valuation such that $v(\alpha\land(\beta\land\gamma))\notin D$.

        \textsc{Case 1:} $v(\alpha\land(\beta\land\gamma))=f$.

        This implies that $v(\alpha)=f$ or $v(\beta\land\gamma)=f$, i.e., $v(\alpha)=f$ or $v(\beta)=f$ or $v(\gamma)=f$. Then, $v(\alpha\land\beta)=f$ or $v(\gamma)=f$. So, $v((\alpha\land\beta)\land\gamma)=f$.
        
        \textsc{Case 2:} $v(\alpha\land(\beta\land\gamma))=n$.
        
        This implies that either $v(\alpha)=n$ and $v(\beta\land\gamma)\neq f$ or $v(\beta\land\gamma)=n$ and $v(\alpha)\neq f$.

        Suppose $v(\alpha)=n$. Then, $v(\alpha\land\beta)=n/f$ and hence $v((\alpha\land\beta)\land\gamma)=n/f$. On the other hand, if $v(\beta\land\gamma)=n$, then either $v(\beta)=n$ and $v(\gamma)\neq f$, or $v(\gamma)=n$ and $v(\beta)\neq f$. In the former situation, $v(\beta)=n$ implies that $v(\alpha\land\beta)=n$ as $v(\alpha)\neq f$. Thus, $v((\alpha\land\beta)\land\gamma)=n$ as $v(\gamma)\neq f$. In the latter situation, where $v(\gamma)=n$, $v((\alpha\land\beta)\land\gamma)=n/f$.
        
        So, in all cases, $v((\alpha\land\beta)\land\gamma))\notin D$. Thus, by contraposition, for any UKN1-valuation $v$, if $v((\alpha\land\beta)\land\gamma)\in D$, then $v(\alpha\land(\beta\land\gamma))\in D$. Hence, $(\alpha\land\beta)\land\gamma)\mone\alpha\land(\beta\land\gamma)$.
    \end{enumerate}

    \begin{enumerate}[label=(\roman*),start=11]
        \item Suppose $v$ is a UKN1-valuation such that $v(\neg\alpha\lor\neg\beta)\notin D$. 

        \textsc{Case 1:} $v(\neg\alpha\lor\neg\beta)=f$.

        Then, $v(\neg\alpha)=v(\neg\beta)=f$. So, $v(\alpha)=v(\beta)=t$, which implies that $v(\alpha\land\beta)=t$. Thus, $v(\neg(\alpha\land\beta))=f$.

        \textsc{Case 2:} $v(\neg\alpha\lor\neg\beta)=n$.

        Then, either $v(\neg\alpha)=n$ and $v(\neg\beta)\neq t$, or $v(\neg\alpha)\neq t$ and $v(\neg\beta)=n$. If $v(\neg\alpha)=n$ and $v(\neg\beta)\neq t$, then $v(\alpha)=n$ and $v(\beta)\neq f$. So, $v(\alpha\land\beta)=n$, and hence $v(\neg(\alpha\land\beta))=n$. The other case when $v(\neg\alpha)\neq t$ and $v(\neg\beta)=n$ similarly leads to $v(\neg(\alpha\land\beta))=n$.

        Thus, in all cases, $v(\neg(\alpha\land\beta))\notin D$. So, by contraposition, for any UKN1-valuation $v$, if $v(\neg(\alpha\land\beta))\in D$, then $v(\neg\alpha\lor\neg\beta)\in D$. Hence, $\neg(\alpha\land\beta)\mone\neg\alpha\lor\neg\beta$.
    \end{enumerate}

    \begin{enumerate}[label=(\roman*),start=14]
        \item Suppose $v$ is a UKN1-valuation such that $v(\neg(\alpha\lor\beta))\notin D$. 

        \textsc{Case 1:} $v(\neg(\alpha\lor\beta))=f$.

        Then, $v(\alpha\lor\beta)=t$, which implies that either $v(\alpha)=t$ or $v(\beta)=t$. So, either $v(\neg\alpha)=f$ or $v(\neg\beta)=f$. This leads to $v(\neg\alpha\land\neg\beta)=f$.

        \textsc{Case 2:} $v(\neg(\alpha\lor\beta))=n$.

        Then, $v(\alpha\lor\beta)=n$, which implies that either $v(\alpha)=n$ and $v(\beta)\neq t$, or $v(\beta)=n$ and $v(\alpha)\neq t$. If $v(\alpha)=n$ and $v(\beta)\neq t$, then $v(\neg\alpha)=n$ and $v(\neg\beta)\neq f$. So, $v(\neg\alpha\land\neg\beta)=n$. The other case, where $v(\beta)=n$ and $v(\alpha)\neq t$ similarly leads to the same conclusion.

        Hence, in all cases, $v(\neg\alpha\land\neg\beta)\notin D$. So, by contraposition, for any UKN1-valuation $v$, $v(\neg(\alpha\lor\beta))\in D$ whenever $v(\neg\alpha\land\neg\beta)\in D$. Thus, $\neg\alpha\land\neg\beta\mone\neg(\alpha\lor\beta)$.
    \end{enumerate}
\end{proof}

Before closing this section, we would like to point out that one could define a material implication in UKN1 as usual: for any $\alpha,\beta\in\lang$, $\alpha\limp\beta:=\neg\alpha\lor\beta$. However, there exists $\alpha,\beta\in\lang$ such that $\{\alpha,\alpha\limp\beta\}\not\mone\beta$, i.e., \emph{modus ponens} fails in UKN1 with respect to material implication. To see this, let $p,q\in\var$ be distinct and $v$ be a UKN1-valuation such that $v(p)=u$ and $v(q)=f$. Then, $v(\neg p\lor q)=t$. Thus, $v(p),v(p\limp q)\in D$ but $v(q)\notin D$.

\section{The logic UKN2}
The logic UKN2 is similar to UKN1 in so far as the signature $\Sigma$, the set of variables $\var$, the set of formulas $\lang$, and the set of truth values $\V=\{t,f,b,n,u\}$ are concerned. The understanding of the states $t,f,b$, and $n$ are as in UKN1, i.e., true, false, both, and not known yet. As for the state $u$, it still represents unknowable but here, in addition to that, it indicates a system error due to which the computer is unable to proceed further with the computation and give any other output. The computer is assumed to come to a halt if it reaches one of the four states $t,f,b$, or $u$, and so these are closed states. On the other hand, $n$ represents that the computation is ongoing and some other output can be expected, which indicates that this is an open state. The closed and non-false values are taken as designated as in the case of UKN1. Thus, $D=\{t,b,u\}$. As for the set of operations corresponding to the connectives, we first note that $f_\neg:\V\to\V$ stays the same as in UKN1. Thus, the following serves as the truth table for negation. 

\[
\begin{array}{c|c|c|c|c|c}
     &t&f&b&n&u\\
     \hline
     f_\neg&f&t&b&n&u
\end{array}
\]

Now, $f_\land,f_\lor:\V^2\to\V$ behave the same way as in UKN1 for the values in $\V\setminus\{u\}$. The state $u$ is infectious here, i.e., if any of the inputs to $f_\land$ or $f_\lor$ is $u$, then the output is also so. This means that when answering a complex question, if an error is encountered during the process of finding the answer to any part, then the answer to the entire question becomes unknowable due to the error. The following are thus the tables for $f_\land$ and $f_\lor$.

\[
\begin{array}{ccc}
  \begin{array}{c|ccccc}
     f_\land&t&f&b&n&u\\
     \hline
     t&t&f&b&n&u\\
     f&f&f&f&f&u\\
     b&b&f&b&n&u\\
     n&n&f&n&n&u\\
     u&u&u&u&u&u
\end{array} &&
\begin{array}{c|ccccc}
     f_\lor&t&f&b&n&u\\
     \hline
     t&t&t&t&t&u\\
     f&t&f&b&n&u\\
     b&t&b&b&n&u\\
     n&t&n&n&n&u\\
     u&u&u&u&u&u
\end{array}
\end{array}
\]
As in the case of UKN1, the values $t,f$, and $b$ behave in the same way as in FDE or FDEe. The value $u$ behaves in the same way as $e$ behaves in FDEe. UKN2, however, differs from FDE and FDEe when it comes to the value $n$. Moreover, unlike FDE or FDEe, UKN2 has three designated values as is the case in UKN1. Nevertheless, since $u$ is infectious over $n$, UKN2 differs from UKN1.

Next, with the matrix $\langle\V,\F,D\rangle$, where $\F$ is the set consisting of $f_\neg,f_\land$, and $f_\lor$ as defined above, we define UKN2-valuations and the UKN2-consequence $\mtwo$ as indicated in general in Section \ref{sec:prelim}. As in the case of UKN1, instead of using the above truth tables, the UKN2-valuations can be alternatively described in a linear form as follows. Suppose $\alpha,\beta\in\lang$ and $v$ is a UKN1-valuation.

\textsc{Negation}
\begin{itemize}
    \item $v(\neg\alpha)=t$ iff $v(\alpha)=f$
    \item $v(\neg\alpha)=f$ iff $v(\alpha)=t$
    \item $v(\neg\alpha)=b$ iff $v(\alpha)=b$
    \item $v(\neg\alpha)=n$ iff $v(\alpha)=n$
    \item $v(\neg\alpha)=u$ iff $v(\alpha)=u$
\end{itemize}

\textsc{Conjunction}
\begin{itemize}
    \item $v(\alpha\land\beta)=t$ iff $v(\alpha)=t$ and $v(\beta)=t$
    \item $v(\alpha\land\beta)=f$ iff ($v(\alpha)=f$ and $v(\beta)\neq u$) or ($v(\beta)=f$ and $v(\alpha)\neq u$)
    \item $v(\alpha\land\beta)=b$ iff ($v(\alpha)=b$ and $v(\beta)=t/b$) or ($v(\beta)=b$ and $v(\alpha)=t/b$) 
    \item $v(\alpha\land\beta)=n$ iff ($v(\alpha)=n$ and $v(\beta)\neq f/u$) or ($v(\beta)=n$ and $v(\alpha)\neq f/u$)
    \item $v(\alpha\land\beta)=u$ iff $v(\alpha)=u$ or $v(\beta)=u$
\end{itemize}

\textsc{Disjunction}
\begin{itemize}
    \item $v(\alpha\lor\beta)=t$ iff ($v(\alpha)=t$ and $v(\beta)\neq u$) or ($v(\beta)=t$ and $v(\alpha)\neq u$)
    \item $v(\alpha\lor\beta)=f$ iff $v(\alpha)=f$ and $v(\beta)=f$
    \item $v(\alpha\lor\beta)=b$ iff ($v(\alpha)=b$ and $v(\beta)=f/b$) or ($v(\beta)=b$ and $v(\alpha)=f/b$) 
    \item $v(\alpha\lor\beta)=n$ iff ($v(\alpha)=n$ and $v(\beta)\neq t/u$) or ($v(\beta)=n$ and $v(\alpha)\neq t/u$)
    \item $v(\alpha\lor\beta)=u$ iff $v(\alpha)=u$ or $v(\beta)=u$
\end{itemize}

We now investigate the properties of the logic UKN2 $=\langle\lang,\mtwo\rangle$. As in the case of UKN1, this logic also doesn't have any tautologies. This can be shown by the same reasoning as earlier. So, once again we have the failure of LEM and LNC, the failure of LEM implying that UKN2 is paracomplete. UKN2 is also paraconsistent, i.e., ECQ fails in UKN2. This can be seen by considering distinct variables $p,q\in\var$ and a UKN2-valuation $v$ such that $v(p)=b$ and $v(q)=n$. Then, $v(\{p,\neg p\})=\{b\}\subseteq D$ but $v(q)\notin D$. Thus, $\{p,\neg p\}\not\mtwo q$. This, along with the failure of LNC, amounts to UKN2 being strongly paraconsistent.

The disjunction introduction rule fails in UKN2 as in the case of UKN1. This can be shown by considering distinct variables $p,q\in\var$ and a UKN2-valuation $v$ such that $v(p)=b\in D$ and $v(q)=n$. Then, $v(p\lor q)=n\notin D$, which implies that $p\not\mtwo p\lor q$. However, the same arguments as in the case of UKN1, show that the standard disjunction elimination rule holds in UKN2. Thus, for any $\alpha,\beta,\gamma\in\lang$, 
\[
\hbox{if }\alpha\mtwo\gamma\hbox{ and }\beta\mtwo\gamma,\hbox{ then }\alpha\lor\beta\mtwo\gamma.
\]

A notable failure of the UKN2-consequence, which is unlike UKN1, is the conjunction elimination rule. There exist $\alpha,\beta\in\lang$ such that $\alpha\land\beta\not\mtwo\alpha$. To prove this, consider distinct variables $p,q\in\var$ and a UKN2-valuation $v$ such that $v(p)=n$ (or $f$) and $v(q)=u$. Then, $v(p\land q)=u$. Thus, $v(p\land q)\in D$, while $v(p)\notin D$, which implies that $p\land q\not\mtwo p$.

However, the conjunction introduction rule holds as usual. It can be seen from the truth table for conjunction that, for any $\alpha,\beta\in\lang$ and for any UKN2-valuation $v$, if $v(\alpha),v(\beta)\in D$, then $v(\alpha\land\beta)\in D$.

The distributive laws fail in UKN2. There exist $\alpha,\beta,\gamma\in\lang$ such that
\begin{enumerate}[label=(\roman*)]
    \item $\alpha\land(\beta\lor\gamma)\not\mtwo(\alpha\land\beta)\lor(\alpha\land\gamma)$; and
    \item $\alpha\lor(\beta\land\gamma)\not\mtwo(\alpha\lor\beta)\land(\alpha\lor\gamma)$.
\end{enumerate}

\begin{proof}
    Same as for UKN1.
\end{proof}

However, for any $\alpha,\beta,\gamma\in\lang$, we have the following entailments.
\begin{enumerate}[label=(\roman*)]
    \item $(\alpha\land\beta)\lor(\alpha\land\gamma)\mtwo\alpha\land(\beta\lor\gamma)$; and
    \item $(\alpha\lor\beta)\land(\alpha\lor\gamma)\mtwo\alpha\lor(\beta\land\gamma)$.
\end{enumerate}

\begin{proof}
    Same as for UKN1 with the extra observation that for any UKN2-valuation $v$ and any formula $\varphi\in\lang$, if $v(\varphi)\notin D$, then in particular, $v(\varphi)\neq u$, which implies that $v(\psi)\neq u$ for any subformula $\psi$ of $\varphi$, and moreover, $v(\theta)\neq u$ for any formula $\theta$ that has such a $\psi$ as a subformula. 
\end{proof}

Thus, we have one-sided distributive laws in UKN2 as in UKN1. We now list some other valid entailments in UKN2. These can be proved using the truth tables. As before, we will prove some as examples. The rest can be argued for similarly. Suppose $\alpha,\beta,\gamma\in\lang$.

\textsc{Double negation laws:}
\begin{multicols}{2}
\begin{enumerate}[label=(\roman*)]
    \item $\neg\neg\alpha\mtwo\alpha$
    \item $\alpha\mtwo\neg\neg\alpha$
\end{enumerate}
\end{multicols}

\textsc{Commutative laws:}
\begin{multicols}{2}
    \begin{enumerate}[label=(\roman*),start=3]
        \item $\alpha\land\beta\mtwo\beta\land\alpha$
        \item $\alpha\lor\beta\mtwo\beta\lor\alpha$
    \end{enumerate}
\end{multicols}

\textsc{Associative laws:}
\begin{multicols}{2}
\begin{enumerate}[label=(\roman*),start=5]
    \item $\alpha\land(\beta\land\gamma)\mtwo(\alpha\land\beta)\land\gamma$
    \item $(\alpha\land\beta)\land\gamma\mtwo\alpha\land(\beta\land\gamma)$
\end{enumerate}
\end{multicols}

\begin{multicols}{2}
\begin{enumerate}[label=(\roman*),start=7]
    \item $\alpha\lor(\beta\lor\gamma)\mtwo(\alpha\lor\beta)\lor\gamma$
    \item $(\alpha\lor\beta)\lor\gamma\mtwo\alpha\lor(\beta\lor\gamma)$
\end{enumerate}
\end{multicols}

\textsc{De Morgan's laws:}
\begin{multicols}{2}
    \begin{enumerate}[label=(\roman*),start=9]
        \item $\neg(\alpha\land\beta)\mtwo\neg\alpha\lor\neg\beta$
        \item $\neg\alpha\lor\neg\beta\mtwo\neg(\alpha\land\beta)$
    \end{enumerate}
\end{multicols}

\begin{multicols}{2}
    \begin{enumerate}[label=(\roman*),start=11]
        \item $\neg(\alpha\lor\beta)\mtwo\neg\alpha\land\neg\beta$
        \item $\neg\alpha\land\neg\beta\mtwo\neg(\alpha\lor\beta)$
    \end{enumerate}
\end{multicols}

\begin{proof}
    The proofs of (i)--(iv) follow straightforwardly from the tables. We prove (vii), (x), and (xi) below. The rest can be proved similarly by analyzing the truth tables and following similar argument patterns.

    \begin{enumerate}[label=(\roman*),start=7]
        \item Suppose $v$ is a UKN2-valuation such that $v((\alpha\lor\beta)\lor\gamma)\notin D$. In particular, this implies that $v((\alpha\lor\beta)\lor\gamma)\neq u$, which means that for any subformula $\psi$ of $(\alpha\lor\beta)\lor\gamma$, $v(\psi)\neq u$, and moreover, $v(\theta)\neq u$ for any formula $\theta$ containing such a $\psi$ as a subformula. In other words, $v(\theta)\neq u$ for any $\theta\in\lang$ that contains a subformula of $(\alpha\lor\beta)\lor\gamma$. Thus, we can safely keep the value $u$ out of all the arguments below.

        \textsc{Case 1:} $v((\alpha\lor\beta)\lor\gamma)=f$.

        Then, $v(\alpha\lor\beta)=v(\gamma)=f$, and $v(\alpha\lor\beta)=f$ implies that $v(\alpha)=v(\beta)=f$. Thus, $v(\beta\lor\gamma)=f$ and this leads to $v(\alpha\lor(\beta\lor\gamma))=f$.

        \textsc{Case 2:} $v((\alpha\lor\beta)\lor\gamma)=n$.

        Then, either $v(\alpha\lor\beta)=n$ and $v(\gamma)\neq t$, or $v(\gamma)=n$ and $v(\alpha\lor\beta)\neq t$. 
        
        \textit{Subcase 1:} $v(\alpha\lor\beta)=n$ and $v(\gamma)\neq t$.
        
        $v(\alpha\lor\beta)=n$ implies that either $v(\alpha)=n$ and $v(\beta)\neq t$, or $v(\beta)=n$ and $v(\alpha)\neq t$. Suppose first that $v(\alpha)=n$ and $v(\beta)\neq t$. Since $v(\gamma)\neq t$ as well, $v(\beta\lor\gamma)\neq t$. Thus, $v(\alpha\lor(\beta\lor\gamma))=n$. In the other case, $v(\beta)=n$, and $v(\gamma)\neq t$, imply that $v(\beta\lor\gamma)=n$. Then, as $v(\alpha)\neq t$ here, $v(\alpha\lor(\beta\lor\gamma))=n$.

        \textit{Subcase 2:} $v(\gamma)=n$ and $v(\alpha\lor\beta)\neq t$.

        $v(\alpha\lor\beta)\neq t$ implies that $v(\alpha)\neq t$ and $v(\beta)\neq t$. So, since $v(\gamma)=n$, $v(\beta\lor\gamma)=n$, and hence, $v(\alpha\lor(\beta\lor\gamma))=n$.

        Thus, in all cases $v(\alpha\lor(\beta\lor\gamma))\notin D$. So, by contraposition, for any UKN2-valuation $v$, if $v(\alpha\lor(\beta\lor\gamma))\in D$, then $v((\alpha\lor\beta)\lor\gamma)\in D$. Hence, $\alpha\lor(\beta\lor\gamma)\mtwo(\alpha\lor\beta)\lor\gamma$.
    \end{enumerate}

    \begin{enumerate}[label=(\roman*),start=10]
        \item Suppose $v$ is a UKN2-valuation such that $v(\neg(\alpha\land\beta))\notin D$. In particular, this implies that $v(\neg(\alpha\land\beta))\neq u$, and thus, by the same arguments as before, $v(\theta)\neq u$ for any formula $\theta$ that contains a subformula of $\neg(\alpha\land\beta)$. We will thus leave the value $u$ of consideration below.

        \textsc{Case 1:} $v(\neg(\alpha\land\beta))=f$.

        Then, $v(\alpha\land\beta)=t$, which implies that $v(\alpha)=v(\beta)=t$. So, $v(\neg\alpha)=v(\neg\beta)=f$. Thus, $v(\neg\alpha\lor\neg\beta)=f$.
        
        \textsc{Case 2:} $v(\neg(\alpha\land\beta))=n$.

        In this case, $v(\alpha\land\beta)=n$, which implies that either $v(\alpha)=n$ and $v(\beta)\neq f$, or $v(\beta)=n$ and $v(\alpha)\neq f$. Then, either $v(\neg\alpha)=n$ and $v(\neg\beta)\neq t$, or $v(\neg\beta)=n$ and $v(\neg\alpha)\neq t$. In either case, $v(\neg\alpha\lor\neg\beta)=n$.
        
        So, in all cases, $v(\neg\alpha\lor\neg\beta)\notin D$. Thus, by contraposition, for any UKN2-valuation $v$, if $v(\neg\alpha\lor\neg\beta)\in D$, then $v(\neg(\alpha\land\beta))\in D$. Hence, $\neg\alpha\lor\neg\beta\mtwo\neg(\alpha\land\beta)$.
    
        \item Suppose $v$ is a UKN2-valuation such that $v(\neg\alpha\land\neg\beta))\notin D$. In particular, this implies that $v(\neg\alpha\land\neg\beta))\neq u$, and thus, by the same reasoning as before, we can keep the value $u$ out of the arguments below.

        \textsc{Case 1:} $v(\neg\alpha\land\neg\beta)=f$.

        Then, either $v(\neg\alpha)=f$, or $v(\neg\beta)=f$, i.e., either $v(\alpha)=t$ or $v(\beta)=t$. This implies that $v(\alpha\lor\beta)=t$, and hence, $v(\neg(\alpha\lor\beta))=f$.

        \textsc{Case 2:} $v(\neg\alpha\land\neg\beta)=n$.

        In this case, either $v(\neg\alpha)=n$ and $v(\neg\beta)\neq f$, or $v(\neg\beta)=n$ and $v(\neg\alpha)\neq f$. So, either $v(\alpha)=n$ and $v(\beta)\neq t$, or $v(\beta)=n$ and $v(\alpha)\neq t$. In both situations, $v(\alpha\lor\beta)=n$, and hence, $v(\neg(\alpha\lor\beta))=n$.

        Thus, in all cases $v(\neg(\alpha\lor\beta))\notin D$. So, by contraposition, for any UKN2-valuation $v$, if $v(\neg(\alpha\lor\beta))\in D$, then $v(\neg\alpha\land\neg\beta)\in D$. Hence, $\neg(\alpha\lor\beta)\mtwo\neg\alpha\land\neg\beta$.
    \end{enumerate}
\end{proof}

As in the case for UKN1, one could define a material implication in UKN2 but modus ponens fails with respect to it. Thus, there exist $\alpha,\beta\in\lang$ such that $\{\alpha,\neg\alpha\lor\beta\}\not\mtwo\beta$. To see this, let $p,q\in\var$ be distinct and $v$ be a UKN2-valuation such that $v(p)=u$ and $v(q)=f$. Then, $v(\neg p\lor q)=u$, and hence, $v(\{p,\neg p\lor q\})\subseteq D$, while $v(q)\notin D$. This failure is, however, not surprising considering the fact that there is an infectious truth value that is designated.

\section{Conclusion}
We have now introduced two new five-valued logics - UKN1 and UKN2, which we call the logics of the unknowable. The five truth values $t,f,b,n$, and $u$ represent, respectively, true, false, both true and false, not known yet, and unknowable. Depending on how the unknowable state is reached, either due to a system error which leads to the stopping of the computation, or through conclusive evidence for it, the fifth value of unknowability becomes infectious or not. The values true ($t$), false ($f$), both ($b$) and unknowable ($u$) are called closed states as the computer halts with these outputs and the answers carry a sense of finality. A value of not known yet ($n$), on the other hand, indicates that the computation is ongoing and there is a possibility of reaching one of the closed states at a future point in time, and hence, is called open. It may be noted that the four-valued reducts of these two logics is the same four-valued logic, which is different from FDE. Thus, UKN1 and UKN2 are distinct from FDEe, the five-valued extension of FDE. The other crucial point of distinction between these and FDEe is the number of designated values. While FDEe treats only the values true and both as designated, the logics discussed in this article have three designated values -- true, both, and the unknowable. These are the closed and non-false values.

We acknowledge that there is another possible five-valued logic that hovers close to the ones discussed here. This would be the same as FDEe but with the fifth value $e$ as designated along with true and both. However, the interpretations of the values would perhaps need to be re-evaluated. We are not sure whether the values, especially $e$ can be interpreted as unknowable. Besides the behavior of the value $n$ is also different in FDEe.

As for future work, the next order of business would be to find sound and complete proof systems for UKN1 and UKN2. The valid entailments listed here can be used to move towards Hilbert systems or natural deduction systems for UKN1 and UKN2. The Soundness and Completeness results can then be worked out.
\bibliography{HCMT}
  \bibliographystyle{plain}
\end{document}